\documentclass[aps,floatfix,twocolumn,showpacs,superscriptaddress,prl]{revtex4}

  \usepackage[dvips]{color,graphicx}
  \usepackage{epsfig}
  \usepackage{bm}
  \usepackage{pstricks}
 
  \usepackage{calc}               % Simple computations with LaTeX variables
  \usepackage{fancybox}           % To have several backgrounds
  \usepackage{fancyhdr}           % Headers and footers definitions
  \usepackage{fancyvrb}           % Fancy verbatim environments
  \usepackage{graphicx}           % Standard graphics package
  \usepackage{multido}            % General loop macro
  \usepackage{pifont}             % Ding symbols (mainly for lists)
  \usepackage{pst-3d}             % PSTricks 3D framed boxes
  \usepackage{pst-grad}           % PSTricks gradient mode
  \usepackage{pst-node}           % PSTricks nodes
  \usepackage{url}                % Convenient URL typesetting
  \usepackage{dcolumn}            % Align decimal points in table

  \def\GeV2{(GeV/c)$^2$}

\begin{document}

\title{Moments of the longitudinal proton structure function
	$\bm{F_{L}}$ \\ from global data
	in the $\bm{Q^{2}}$ range 0.75~--~45.0~(GeV/c)$\bm{^{2}}$} 

\author{P.~Monaghan}
\affiliation{Hampton University, Hampton, VA 23668, USA}
\author{A.~Accardi}
\affiliation{Hampton University, Hampton, VA 23668, USA}
\affiliation{Jefferson Lab, Newport News, VA 23606, USA}
\author{M.~E.~Christy}
\affiliation{Hampton University, Hampton, VA 23668, USA}
\author{C.~E.~Keppel}
\affiliation{Hampton University, Hampton, VA 23668, USA}
\author{W.~Melnitchouk}
\affiliation{Jefferson Lab, Newport News, VA 23606, USA}
\author{L.~Zhu}
\affiliation{Hampton University, Hampton, VA 23668, USA}

\begin{abstract}
We present an extraction of the lowest three moments of the proton
longitudinal structure function $F_L$ from world data between
$Q^2 = 0.75$ and 45~(GeV/c)$^2$.  The availability of new $F_L$ data
at low Bjorken $x$ from HERA and at large $x$ from Jefferson Lab allows
the first determination of these moments over a large $Q^2$ range,
relatively free from uncertainties associated with extrapolations
into unmeasured regions. The moments are found to be underestimated by
leading twist structure function parameterizations, especially for the
higher moments, suggesting either the presence of significant higher
twist effects in $F_L$ and/or a larger gluon distribution at high $x$.
\end{abstract}

\date{\today}


\maketitle

\bibliographystyle{apsrev}

%%%%%%%%%%%%%%%%%%%%%%%%%%%%%%%%%%%%%%%%%%%%%%%%%%%%%%%%%%%%%%%%%%%%%%%%%
{\it Introduction.}\ \ 
The suppression of the longitudinal deep-inelastic lepton--proton
scattering cross section relative to the transverse cross section
was an important early verification of the quarks' spin-1/2 nature.
In fact, for a point-like quark the longitudinal structure function
$F_L$ is identically zero.  For a composite particle such as a proton,
$F_L$ is small but finite, and its exact value and momentum dependence
reflect the quantum interaction effects between the proton's quark
and gluon (or parton) constituents.

In QCD, one of the novel features of the proton longitudinal structure
function is its strong sensitivity to the nonperturbative initial
state distribution of gluons.  The moments of $F_L$ in particular are
related to matrix elements of local twist-two operators, which can be
computed directly in lattice QCD.  Traditionally, the gluon
distribution $g(x)$ has been largely determined by studying the $Q^2$
evolution of the $F_2$ structure function, which at high photon
virtualities is dominated by the transverse cross section.
In recent global fits
\cite{Martin:2009iq, Alekhin:2009ni, Accardi:2011fa, Ball:2011uy,
JimenezDelgado:2008hf} $g(x)$ is further constrained at low parton
momentum fraction $x$ by jet production data in hadronic collisions.

At large values of $x$, where the cross sections are small, the
extraction of the gluon density becomes increasingly difficult,
leading to large uncertainties in $g(x)$ at $x \gtrsim 0.3$.
As a result, the higher moments of $F_L$, which are weighted
towards higher values of $x$, are particularly challenging.

Data on $F_L$ are generally difficult to extract from cross section
measurements, requiring detailed longitudinal-transverse (L/T)
separations in which experiments are performed at the same $x$ and
$Q^2$ but at different energy.  Historically, the kinematic range
spanned by $F_L$ data was therefore rather limited, typically
concentrated in the small- and intermediate-$x$ regions, whereas a
precise moment analysis necessitates a broad coverage in $x$ at fixed
$Q^2$.  Previous moment analyses consequently required recourse to
model-dependent estimates of the longitudinal to transverse structure
function ratio \cite{Ricco:1998yr}, rendering a precise evaluation of
$F_L$ moments and their uncertainties problematic.

Recently, new data on the proton longitudinal structure function have
been taken at low $x$ from the H1 experiment \cite{Collaboration:2010ry}
at HERA, and at large $x$ from Hall C at Jefferson Lab
\cite{Liang:2004tj}.  The latter in particular cover a significant
$x$ range from $Q^2 = 4$~(GeV/c)$^2$ down to $Q^2 < 1$~(GeV/c)$^2$.
Combined with the previous $F_L$ results, the new data allow for the
first time a direct extraction of the $Q^2$ dependence of several low
$F_L$ moments over a large range of $Q^2$.

In this Letter, we report the results of such an extraction, with an
accurate determination of the lowest three moments of $F_L$, together
with their uncertainties.  Comparison of these to moments computed
from parametrizations of leading twist parton distribution functions
(PDFs) \cite{Martin:2009iq, Alekhin:2009ni, Accardi:2011fa} then
allows one to draw conclusions about the (poorly constrained) gluon
distribution at large $x$, or the role of higher twist effects in the
longitudinal cross section.

%%%%%%%%%%%%%%%%%%%%%%%%%%%%%%%%%%%%%%%%%%%%%%%%%%%%%%%%%%%%%%%%%%%%%%%%%
{\it Data Analysis.}\  \
Our analysis is performed in terms of the longitudinal Nachtmann
moments, defined as \cite{Nachtmann:1973mr}
\begin{eqnarray}
\label{eq:NachtLMom}
M_L^{(n)}(Q^2)
= \int_0^1 dx\ { \xi^{n+1} \over x^3 }
    \Big\{  F_L(x,Q^2) \hspace{2.5cm}	\\   
	 +\ 2(\rho^2-1){ (n+1)/(1+\rho) - (n+2) \over (n+2)(n+3) }
	    F_2(x,Q^2)
    \Big\},				\nonumber
\end{eqnarray}
where the Nachtmann scaling variable $\xi = 2x/(1+\rho)$,
with $\rho = \sqrt{1+4 M^2 x^2/Q^2}$, and $M$ is the proton mass
\cite{Greenberg:1972es,Nachtmann:1973mr}.  The Nachtmann moments are
constructed to remove from the data the kinematic dependence on the
target mass \cite{Georgi:1976ve,Schienbein:2007gr,Brady:2011uy},
thus allowing a direct comparison with the Cornwall-Norton moments
calculated from leading twist PDFs.
At very low $Q^2$ this treatment may generate some residual uncertainty
if the contributions from the threshold region at $\xi \to 1$ are large
\cite{Steffens:2006ds}.

\begin{figure}[t]
\vskip-.5cm
\includegraphics[width=\linewidth]{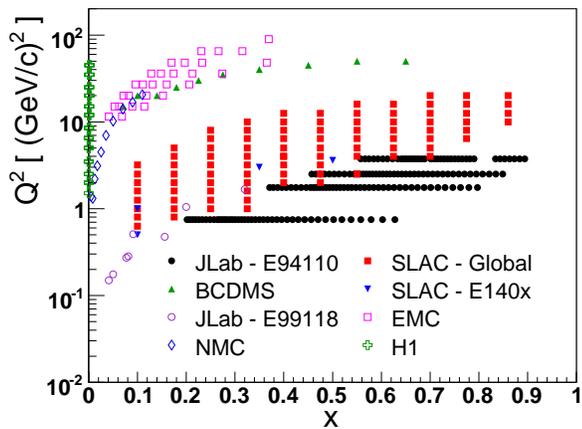}
\caption{(Color online)
	The $Q^2$ and $x$ distribution of $F_L$ data sets used in this
	analysis. The new H1 data \cite{Collaboration:2010ry} at very
	small $x$ appear clustered around the vertical axis because
	of the linear $x$ scale.}
\label{fig:dataqx1}
\vspace{-0.5cm}
\end{figure}

In this analysis only $F_L$ values extracted from dedicated,
experimental L/T separations of the proton cross section data are used.
This constraint is a critical requirement to avoid the introduction
of model dependence into the $M_L^{(n)}$ extraction, and to accurately
estimate the uncertainties on the moments.
The utilized proton $F_L$ data come from a range of experiments at
CERN (EMC \cite{Aubert:1985fx},
      BCDMS \cite{Benvenuti:1989fm},
      NMC \cite{Arneodo:1996qe}),
SLAC (E140X \cite{re140x},
      SLAC global \cite{Whitlow:1991uw}),
DESY (H1 \cite{Collaboration:2010ry}), and
Jefferson Lab (E94110 \cite{Liang:2004tj},
	       E99118 \cite{PhysRevLett.98.142301}).
The regions of the $Q^2$--$x$ space covered by the data sets are
shown in Fig.~\ref{fig:dataqx1}.

\begin{figure}[t]
\includegraphics[width=\linewidth]{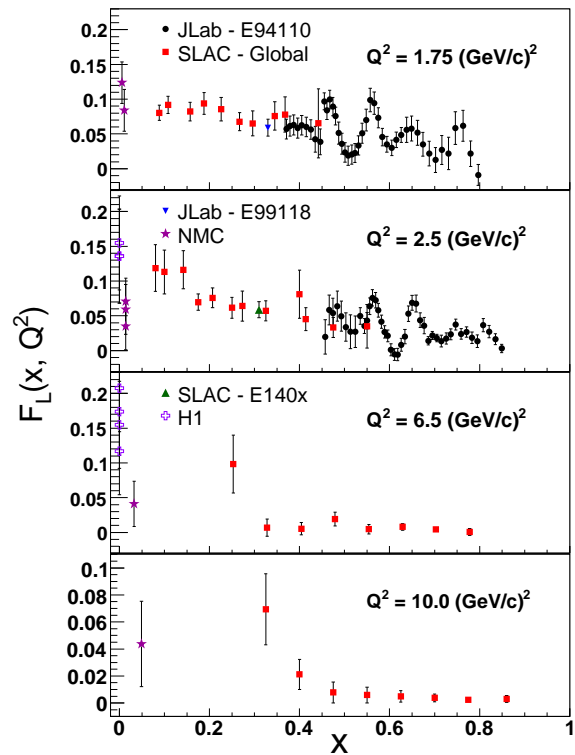}
\caption{(Color online)
	Example plots of the $F_L$ data used in this analysis for
	several $Q^2$ bins in the range $1.75 - 10.0$~(GeV/c)$^2$.}
\label{fig:flstack}
\vspace{-0.5cm}
\end{figure}

Since much of the data were not taken at fixed $Q^2$ values,
the $Q^2$ bins were chosen to ensure the broad coverage in $x$
necessary for a moment extraction.  For instance, a typical bin
at $Q^2 = 6.5$~(GeV/c)$^2$ included all data in the range of
$6 < Q^2 < 7$~(GeV/c)$^2$.  To account for any $Q^2$ dependence
in the bin, the data were bin-centered to the central $Q^2$
utilizing a combination of global data fits which gave good
descriptions of the data over the relevant kinematic range.
A sample of these data bin-centered in $Q^2$ is shown for
several central $Q^2$ values in Fig.~\ref{fig:flstack}.
For the $x$ integration, the data were then binned in $x$
from 0.01 to 1, utilizing bins of width $\Delta x=0.01$.

The data shown in Fig.~\ref{fig:flstack} provide the most
comprehensive kinematical coverage of $F_L$ to date.  However, some
regions of $x$ with sparse data remain, especially at larger $Q^2$.
These gaps were filled by utilizing phenomenological fits to
calculate the structure function at the center of any empty $x$ bin.
For data with $W^2 > 3.9$~(GeV/c)$^2$, a model obtained by a fit
\cite{Christy:FLfit} to world data was used, while for
$W^2 < 3.9$~(GeV/c)$^2$ a fit to the resonance region data was
employed \cite{liang-phd}.
(These models were also used for the bin-centering in $Q^2$,
discussed above.)

The fit values of $F_L$ were also renormalized bin-by-bin.
For $x > 0.4$, where experimental data are abundant, the fit value
was renormalized by the ratio of the fit to the error-weighted mean
of the real data points at the start and end of each empty interval
in $x$.  For intervals with $x<0.4$, where the $x$ gaps are large
and data scarce, the model was renormalized using the error-weighted
mean of all real data points up to $x = 0.4$, to prevent a single
data point at low $x$ from biasing the renormalization factor.

Since the longitudinal Nachtmann moment defined in
Eq.~(\ref{eq:NachtLMom}) includes contributions from both $F_L$ and
$F_2$, the entire analysis was repeated for the $F_2$ data of the same
experiments. After filling the gaps in $x$ for each $Q^2$ bin, each
structure function was integrated as in Eq.~(\ref{eq:NachtLMom}) and
the separate contributions were summed together to generate the
longitudinal Nachtmann moments $M_L^{(n)}$ for $n=2$, 4 and 6.  The
limits of integration were taken from $x = 0.01$ up to the inelastic
or pion production threshold, $x_{\pi}=[1 + (m_\pi^2 +
2Mm_\pi)/Q^2]^{-1}$, with $m_\pi$ the pion mass. Above this point,
which is below $x=1$, only elastic scattering is possible in the
laboratory. Results with and without the elastic contribution included
are presented below.

A Monte-Carlo procedure was utilized to evaluate the errors on the
moments.  For this purpose, 1000 pseudo-data sets were generated for
each $Q^2$ bin and for each structure function, by sampling for each
data point a Gaussian distribution with mean value and width equal to
the value and total error of the data point.  For each pseudo-data
set, the $x$ coverage gaps were filled using the method described
above, and the data integrated to obtain pseudo-moments. This resulted
in two distributions, one each for the $F_L$ and $F_2$
contributions to the moment. The value of the Nachtmann moment was
defined as the sum of the mean value of the pseudo-moment
distributions for the $F_L$ and $F_2$ components. The statistical
error was defined as the sum in quadrature of the standard deviations
of each of the two pseudo-moment distributions.

In order to estimate a model dependent (systematic) error, the process
of filling in the gaps in the data was repeated for three other
combinations of models.  The first two used the ALLM parameterization
for $F_2$ \cite{Abramowicz:1997ms} and the R1990 model for the
longitudinal to transverse ratio \cite{Whitlow:1990dr} in the region
with $W^2 > 3.9$~(GeV/c)$^2$, in combination with two resonance region
fits \cite{Christy:2007ve,liang-phd}.  The third combination used the
fit to world data from Ref.~\cite{Christy:FLfit} for
$W^2 > 3.9$~(GeV/c)$^2$ and the resonance region \cite{Christy:2007ve}
for lower $W$.  After generating distributions of pseudo-moments for
all model combinations, the systematic error was defined as the maximum
difference between the original model combination used to fill in the
gaps and any of the other three combinations.  The systematic error
only has a significant contribution to the total error for the first
two $Q^2$ bins of the $n=2$ moment; otherwise, the statistical error
dominates.  Finally, the total error on each data point was calculated
as the sum in quadrature of the statistical and systematic errors.

%%%%%%%%%%%%%%%%%%%%%%%%%%%%%%%%%%%%%%%%%%%%%%%%%%%%%%%%%%%%%%%%%%%%%%%%%
{\it Results.}\  \
The extracted $n=2$, 4 and 6 longitudinal Nachtmann moments are given
in Table~\ref{mlresults}, for $Q^2$ between 0.75 and 45~$\rm
(GeV/c)^2$.  The values include the measured inelastic contributions
as well as the elastic component, computed from the global proton form
factor parameterization in Ref.~\cite{Arrington:2007ux}.  The elastic
contribution is significant only for the lowest $Q^2$ bins, and
decreases rapidly for $Q^2 \gtrsim 2$~\GeV2.  The errors are largely
driven by the uncertainty on the $F_L$ data.

\newcolumntype{d}[1]{D{.}{.}{#1}}

\begin{table}[!htb]
\begin{center}
%  \begin{tabular}{|c|cc|cc|cc|} 
   \begin{tabular}{|d{5}|d{1}l|d{1}l|d{2}l|}
   \hline 
    \multicolumn{1}{|c|}{$Q^{2}$  } & \multicolumn{6}{c|}{$M_{L}^{(n)}\ \times 10^{-3}$} \\
%     \cline{2-7} 
    \multicolumn{1}{|c|}{(GeV/c)$^{2}$  } & \multicolumn{2}{c}{~~$n = 2$} & \multicolumn{2}{c}{~$n = 4$} & \multicolumn{2}{c|}{~$n = 6$~} \\ 
    \hline
    ~~~~0.75(\rm el) & ~~12.4 & $\pm$ 5.7 &  ~~8.3 & $\pm$ 1.9 & ~~3.64 & $\pm$ 0.63~  \\
    0.75      & 19.7 & $\pm$ 3.3 &  1.2 & $\pm$ 0.4 & 0.18 & $\pm$ 0.08  \\
    1.75(\rm el) & -1.6 & $\pm$ 0.6 & \multicolumn{1}{r}{~0.09} & $\pm$ 0.28 & 0.31 & $\pm$ 0.13  \\
    1.75  & 29.7 & $\pm$ 2.4 &  2.8 & $\pm$ 0.2 & 0.60 & $\pm$ 0.07  \\
    2.5(\rm el) & -1.0 & $\pm$ 0.2 & \multicolumn{1}{l}{~$-$0.23} & $\pm$ 0.09 & -0.03 &$\pm$ 0.05 \\
    2.5  & 27.0 & $\pm$ 4.7 &  2.9 & $\pm$ 0.3 & 0.76 & $\pm$ 0.07  \\
    3.75(\rm el) & -0.4 & $\pm$ 0.05 & \multicolumn{1}{l}{~$-$0.14} & $\pm$ 0.02 &-0.06 & $\pm$ 0.01 \\
    3.75 & 17.5 & $\pm$ 7.7 &  1.6 & $\pm$ 0.4 & 0.46 & $\pm$ 0.08  \\
    5.0  & 16.3 & $\pm$ 6.7 &  1.0 & $\pm$ 0.7 & 0.26 & $\pm$ 0.30 \\
    6.5  & 7.7 & $\pm$ 6.3  &  0.6 & $\pm$ 0.4 & 0.17 & $\pm$ 0.12 \\
    8.0  & 24.7 & $\pm$ 14.1 &  1.3 & $\pm$ 0.5 & 0.26 & $\pm$ 0.11 \\
    10.0 & 15.5 & $\pm$ 8.8 &  0.9 & $\pm$ 0.4 & 0.20 & $\pm$ 0.10 \\
    15.0 & 2.7 & $\pm$ 9.8  &  0.4 & $\pm$ 0.6 & 0.13 & $\pm$ 0.09  \\
    20.0 & 0.2 & $\pm$ 9.1 &  -0.2 & $\pm$ 0.8 & 0.01 & $\pm$ 0.14  \\
    45.0 & 5.4 & $\pm$ 11.0~  &  0.2 & $\pm$ 0.3 & 0.0 & $\pm$ 0.05 \\ 
    \hline
  \end{tabular}
  \caption{The experimental longitudinal Nachtmann moments $M^{(n)}_{L}$ 
(scaled by $10^{-3}$) extracted from the data along with their statistical 
errors. The inelastic results for the $n=2$, 4 and 6 moments are given for 
each $Q^{2}$ bin, with the elastic contribution shown only for the four 
lowest $Q^{2}$ bins.}
  \label{mlresults}
\end{center}
\end{table}
% \vspace{-0.5cm}

\begin{figure}
\includegraphics[width=\linewidth]{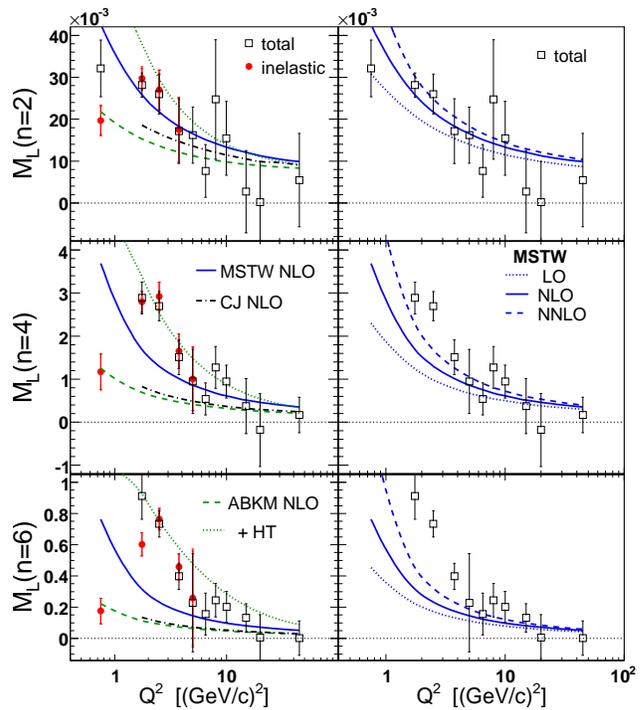}
\caption{(Color online)
	Longitudinal Nachtmann moments $M_L^{(n)}$ for
	$n=2, 4$ and 6, as a function of $Q^2$, for the measured
	inelastic (circles) and inelastic plus elastic (squares)
	contributions.  The left panels compare the data with
	calculations based on global NLO fits to PDFs from MSTW08
	\cite{Martin:2009iq}, ABKM09 \cite{Alekhin:2009ni} (including
	also HT corrections) and CJ \cite{Accardi:2011fa}, while the
	right panels show the comparison with the MSTW08 fits for
	different orders in $\alpha_s$ (LO, NLO and NNLO).}
\label{fig:mln2}
\vspace{-0.5cm}
\end{figure}

The experimental Nachtmann moments are shown in Fig.~\ref{fig:mln2},
and compared with calculations of the Cornwall-Norton moments of
$F_L$ using the MSTW08 \cite{Martin:2009iq}, ABKM09 \cite{Alekhin:2009ni}
and CTEQ-Jefferson Lab (CJ) \cite{Accardi:2011fa} global PDF
parameterizations.

The MSTW08 fit included data on the $F_2$ and $F_L$ structure functions
in fixed target experiments and HERA collider data on reduced DIS
cross sections satisfying $Q^2>2$~\GeV2 and $W^2>15$~\GeV2.
The kinematic cuts were imposed to avoid the region where higher twist
(HT) effects may be significant, thereby excluding data at high $x$.
Nuclear corrections in deuterium DIS were not included in the fit,
although these strongly affect the $d$ quark and gluon PDFs, even within
the cuts used \cite{Accardi:2009br}.  Jet data in $pp$ collisions were
also included, constraining the gluon PDF at $x \lesssim 0.3$.

For the CJ analysis \cite{Accardi:2011fa}, a similar data set to that
used by MSTW08 was fitted, although $F_L$ data were not directly included.
The primary constraint here on the $F_L$ structure function, and the
large-$x$ gluon PDF, was therefore from scaling violations in the
$F_2$ data.  Significantly, the kinematical cuts were relaxed to
$Q^2 > 1.69$~(GeV/c)$^2$ and $W^2 > 3$~(GeV/c)$^2$, increasing
considerably the large-$x$ coverage afforded by the high-precision
SLAC and Jefferson Lab data.  The less restrictive cuts necessitated
inclusion of target mass (TM) and HT contributions, and nuclear
corrections were incorporated using several models of the deuteron.
Since the DIS data were limited to the $F_2$ structure function,
only the leading twist contribution to $M_L^{(n)}$ could be computed
from these PDFs.

The ABKM09 fit \cite{Alekhin:2009ni} used similar cuts to the CJ analysis,
and also included nuclear, TM and HT corrections, but did not utilize
Jefferson Lab data.  Fits were performed to DIS cross sections directly,
rather than to the extracted $F_2$, although the exclusion of jet data
from the analysis weakens the constraints on the gluon PDF at
$x \lesssim 0.3$.  Furthermore, HT terms were included for both
$F_2$ and $F_L$, allowing calculation of moments up to twist 4.

Comparison of the measured moments with the PDF-based NLO calculations
in Fig.~\ref{fig:mln2}~(left) shows a turnover of the inelastic
moments at $Q^2 \approx 2-3$~$\rm (GeV/c)^2$. This is due to the
effect of the pion threshold $x_{\pi}$, which decreases the limit of
integration of the real data at low $Q^2$ values. At about the same
$Q^2$ value, the elastic contribution also becomes non-negligible.
Theoretical calculations of the DIS cross sections which neglect the
pion threshold and integrate up to $x=1$ can therefore be meaningfully
compared to data only for $Q^2\gtrsim 3$~$\rm (GeV/c)^2$.

The leading twist calculations are in generally good agreement with
the data for $Q^2 \gtrsim 10$~$\rm (GeV/c)^2$.  At lower $Q^2$ the
fits underestimate the data, particularly for the higher moments where
large $x$ plays an increasing important role.  The disagreement with
the low-$Q^2$, large-$n$ data may reflect the poorly constrained gluon
PDF, or possibly large effects from higher order perturbative QCD or
or higher twist corrections at large values of $x$.

The effect of including higher order terms is illustrated in
Fig.~\ref{fig:mln2}~(right), where the total Nachtmann moments are
compared with moments calculated from the MSTW08 PDFs
\cite{Martin:2009iq} at leading order (LO), next-to-leading order
(NLO) and next-to-next-to-leading order (NNLO).  While the LO results
generally underestimate the data at low $Q^2$, the agreement
progressively improves with increasing order.  It is only in the
highest ($n=6$) moment that any discrepancy appears, possibly
indicating some underestimation of $g(x)$ at high $x$.
There is a well known and large uncertainty on $g(x)$ which can also
be observed in the substantial differences in the NLO calculations
from different PDFs, as shown in Fig.~\ref{fig:mln2}~(left).

The role of HT contributions in explaining the missing strength
at small $Q^2$ can be explored by comparing the ABKM09 fit
\cite{Alekhin:2009ni} with and without higher twist contributions.
Inclusion of HT corrections improves the agreement with data,
but overestimates the strength somewhat, even within the relatively
large uncertainty of the HT contributions.  This remains true even in
the ABKM09 NNLO fit (not shown), where the LT contribution increases,
albeit more slowly than in the MSTW08 fit, and the HT terms decrease,
leaving the total curve stable.  Since the ABKM09 fit does not include
the recent Jefferson Lab data \cite{PhysRevC.70.015206,
PhysRevLett.98.142301}, it is not directly constrained at lower $Q^2$
and larger $x$.  Minimizing extrapolation uncertainties and precisely
studying the interplay of leading and higher twist contributions in
this region will require use of the new data in global fits.

%%%%%%%%%%%%%%%%%%%%%%%%%%%%%%%%%%%%%%%%%%%%%%%%%%%%%%%%%%%%%%%%%%%%%%%%%
{\it Conclusions.}\  \
In summary, we have extracted the lowest three Nachtmann moments
of the proton longitudinal structure function over the range
$Q^2=0.75-45.0$~$\rm (GeV/c)^2$ from world $F_L$ data augmented by
recent small-$x$ measurements by the H1 collaboration at HERA and
large-$x$ measurements from Jefferson Lab.  Although the data coverage
in $x$ is not absolute, four different combinations of models have
been used to fill gaps in the data, allowing the moments to be
calculated with a rigorous uncertainty analysis.  A reasonable
estimate of the errors has been obtained by a statistical analysis
of the moments calculated from a Monte-Carlo procedure for sampling
the data.

Comparison of the experimental moments with perturbative QCD
calculations using recent global PDF fits reveals a need for either
increased high-$x$ gluon distribution, which is possible given its
large uncertainty, or possibly non-negligible higher twist effects.
Discrepancies between data and the PDFs increase as the moment order
increases, but are reduced by increasing the order of perturbation
theory.  Neglecting higher twist terms or higher order calculations in
the global fits may overestimate the extracted gluon distributions at
large $x$, as shown by the comparison of ABKM09 and MSTW08 calculations.

Relatively good constraints on the large-$x$ gluon can already be
obtained from the scaling violations of $F_2$, provided that a weak
cut on $W$ is considered along with TM and HT included, as shown
in the CJ fits \cite{Accardi:2011fa}.  However, inclusion of $F_L$
data in global fits is required to provide direct constraints on the
gluons, with the recent Jefferson Lab data at low $Q^2<4$~GeV$^2$,
in particular, facilitating a considerable extension of reach in $x$.

%%%%%%%%%%%%%%%%%%%%%%%%%%%%%%%%%%%%%%%%%%%%%%%%%%%%%%%%%%%%%%%%%%%%%%%%%
%% \acknowledgements

We thank A.~Psaker for his collaboration in the early stages of this
work, and R.~Ent for many informative discussions.
This work was supported by the US Department of Energy contract
No.~DE-AC05-06OR23177, under which Jefferson Science Associates, LLC
operates Jefferson Lab, and US National Science Foundation award
No.~1002644.

%%%%%%%%%%%%%%%%%%%%%%%%%%%%%%%%%%%%%%%%%%%%%%%%%%%%%%%%%%%%%%%%%%%%%%%%%
\bibliography{references}

\end{document}